\documentclass[showpacs,prb,letterpaper,twocolumn,floatfix,nobalancelastpage]{revtex4}

\usepackage{hyperref}
\hypersetup{colorlinks}
\usepackage{amsmath}
\usepackage{graphicx}
\usepackage{color}
\usepackage{calc} 
\usepackage{xspace} 
\usepackage{todonotes}
\usepackage{overpic}

\newcommand{\nm}{\ensuremath{\mathrm{\,nm}}\xspace}

\newcommand{\G}[1][{}]{\ensuremath{{\overline{\mathbf G}}_\mathrm{#1}}}
\newcommand{\e}{\mathbf e}
\newcommand{\id}{\mathrm d}
\newcommand{\R}{\mathbf r}

\newcommand{\zerofoot}[1]{{\lefteqn{\scriptstyle{#1}}}\ }

\begin{document}
\title{Interplay between disorder and local field effects in photonic crystal waveguides}
\author{M. Patterson}
\author{S. Hughes}
\email{shughes@physics.queensu.ca}

\date{\today}

\begin{abstract}
We introduce a theory to describe disorder-induced scattering in photonic crystal waveguides, specifically addressing the influence of local field effects and scattering within high-index-contrast perturbations. Local field effects are shown to increase the predicted disorder-induced scattering loss and result in significant resonance shifts of the waveguide mode. We demonstrate that two types of frequency shifts can be expected, a mean frequency shift and a RMS frequency shift, both acting in concert to blueshift and broaden the nominal band structure. For a representative waveguide, we predict substantial meV frequency shifts and band structure broadening for a telecommunications operating frequency, even for state of the art fabrication. The disorder-induced broadening is found to increase as the propagation frequency approaches the slow light regime (mode edge) due to restructuring of the electric field distribution. These findings have a dramatic impact on high-index-contrast nanoscale waveguides, and, for  photonic crystal waveguides, suggest that the nominal slow-light mode edge may not even exist. Furthermore, our results shed new light on why it has hitherto been impossible to observe the very slow light regime for photonic crystal waveguides.
\end{abstract}

\maketitle

\begin{figure}[t!]
	\centering
	\begin{overpic}[]{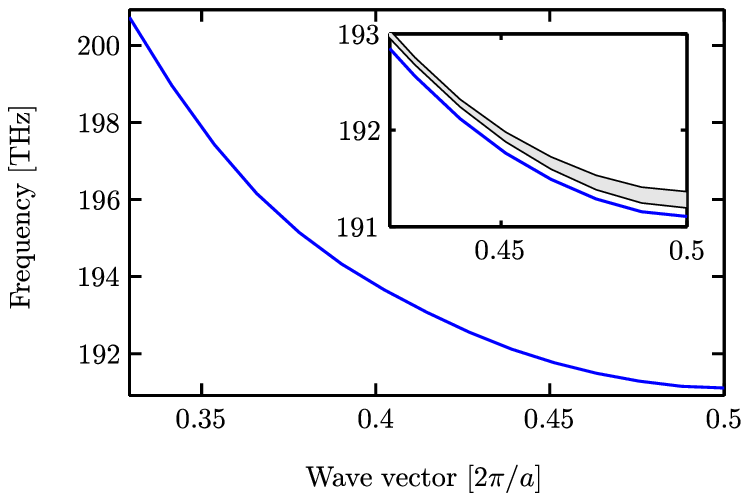}\put(-4,65){a)}\end{overpic}
	\vspace{0.3cm}
	\vspace{0.1cm}
	\begin{overpic}[]{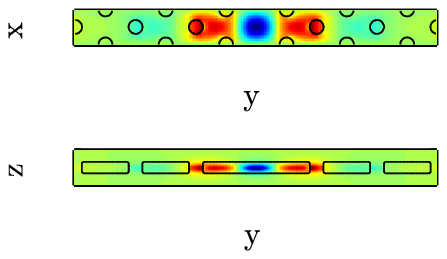}\put(0,55){b)}\end{overpic}
\hspace{0.5cm}
	\begin{overpic}[width=1.1in]{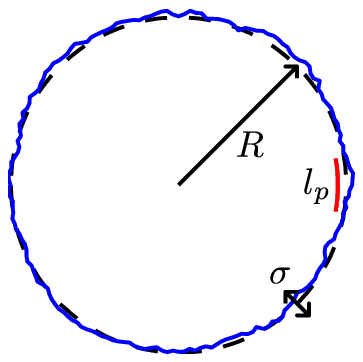}\put(-10,72){c)}\end{overpic}
	\caption{\label{fig:bandstructure} \label{fig:RoughHole} (Color online) (a) Nominal dispersion of PC waveguide mode (blue), and the broadened disorder-induced band structure that we will introduce later (grey shading).
 (b) Electric field Bloch mode near the band edge along symmetry planes in the unit cell. (c) Example of a disordered hole profile that
 satisfies the statistics used in the calculation (blue). The ideal radius (dashed black) and correlation length (short red arc) are shown for reference,
where $R$ indicates the nominal radius of the unperturbed hole and $\sigma$ is the
RMS fluctuation.}
\end{figure}

Photonic crystal (PC) structures comprised of high-index-contrast cavities and waveguides offer a rich degree of control over light-matter interactions, leading to trapped \cite{Akahane:2003} and slow light modes \cite{Baba:2008a} buried within a forbidden photonic band gap. In a planar PC semiconductor system,  waveguide modes can be completely bound below the light line. However,  {manufacturing imperfections} result in
{\em fabrication disorder} that breaks the translational invariance of a nominally perfect lattice, and causes external scattering of the bound modes. From a theoretical perspective, the role and impact of fabrication disorder on the effect of slow-light slab waveguides is becoming better understood  \cite{Hughes:2005, Povinelli:2004, Gerace:2004}; minute sidewall imperfections act to perturb the propagating mode causing the light to out scatter through radiation modes or backscatter and multiple scatter within the waveguide \cite{Patterson:2009b,Patterson:2009,Mazoyer:2009}. Scattering losses become particularly pronounced in the slow light regime as the local density of states of the mode into which light can scatter increases. Although many experiments have confirmed this slow light loss behaviour \cite{Kuramochi:2005,OFaolain:2007,Engelen:2008}, an open question that still remains is what is the effect of disorder on the band structure? The answer to this question involves a complex interplay between local field corrections and enforcing boundary conditions between parallel and perpendicular field components. Experimental evidence for a dramatic reduction of transmission near the mode edge is observed to take place much sooner than that predicted for the scaling of scattering loss, e.g., see Ref.~\citenum{Patterson:2009b}. This suggests that some unknown effect is either shifting the mode edge or broadening the band structure (see Fig.~\ref{fig:bandstructure}(a)). Here we address this question directly and show that surprisingly large changes in the band structure occur which have not been anticipated before. Our findings  introduce a dramatic revision of our present understanding of the role of disorder-induced scattering in these intriguing
nanostructures.

The usual theoretical approaches to modelling disorder-induced scattering, employ standard perturbation theory where the nominal ({\em disorder free}) electric field is used with the dielectric index change to model polarization scatterers, through ${\bf P}_{\rm dis} = \Delta \varepsilon {\bf E}_0$, with ${\bf E}_0$  the unperturbed field and $\Delta \varepsilon$ is the dielectric contrast change resulting from a spatial perturbation (e.g., air to silicon gives $\Delta \varepsilon\approx 12$) \cite{note1}. For high-index-contrast perturbations, as shown by Johnson {\em et al.}~\cite{Johnson:2005}, this polarization scatterer is problematic for several reasons: (i) the parallel components of the electric field and the perpendicular components of the displacement field must be continuous across the surface, and (ii) the index change results in local field corrections. Andreani and Gerace\cite{Andreani:2007} attempted to estimate the magnitude of this error in their theory by comparing with a
 simple numerical supercell calculation of perfect hole shapes with different
 radii, and they concluded that the effects of local fields was not important. Wang {\em et al.} \cite{Wang:2008} employed 
 the 
 well known {\em slowly-varying surface approximation}~\cite{Johnson:2005}, valid {\em only} for smooth bumps, and demonstrated the impact on increasing the scattering losses. For PC cavity systems, Ramunno and Hughes \cite{Ramunno:2009} showed that a {\em quickly-varying surface perturbation} can cause mean resonance shifts of a strongly confined cavity mode \cite{Ramunno:2009}. RMS ensemble average shifts and band edge broadening have also been predicted (without including local field effects) for intrinsically lossy coupled-cavity PC waveguides \cite{Fussell:2008,Mookherjea:2007}.

%
%

%
%
With regards to PC waveguides, to the authors' knowledge, there have been no calculations nor any awareness of disorder-induced resonance shifts. Neither has anyone computed disorder-induced losses in the presence of realistic rapidly-varying surface perturbation by properly addressing the two problematic criteria above. In this work, we overcome this limitation, and introduce a  straightforward optical scattering theory that allows one to compute  scattering losses and resonance shifts within an ensemble averaging procedure, while accounting for local field effects. The {\em ensemble} average statistical calculation applies to measurements of an ensemble
of nominally identical structures, and (or) it applies to the statistics of one waveguide
which contains many nominally identical unit cells. This is a reasonable first model to report, since typical PC waveguides of only 1\,mm-length contain already many thousands of unit cells; our methodology can also help guide future coherent scattering theories, where such effects were neglected \cite{Patterson:2009}. Using representative calculations, we subsequently demonstrate that significant and unusual disorder-induced changes in the band structure can occur, as well as significant modifications (increases) to the predicted scattering losses. These predictions are interesting in their own right and are important for the analysis and interpretation of related  experiments.

The total electric field, in the presence of disorder, can be calculated from an integral solution of the Maxwell equations, through
\begin{align}
%
	\mathbf E(\R;\omega) 
&= \mathbf E_i(\R;\omega) + 
\int_\zerofoot{\mathrm{all\ space}} 
 \id\R' \,\G(\R,\R';\omega) \cdot \mathbf P(\R';\omega),
	\label{eqn:dyson}
\end{align}
where $\mathbf P(\R';\omega)$ is the polarization-like density due to the disorder in the system, $\mathbf E_{i}(\R;\omega)$ is the electric field in the ideal (no disorder) system\cite{note1}, and $\G(\R,\R';\omega)$ is the photon Green function. The Green function is simply a dipole solution to the Maxwell wave equation and it contains information about how light scatters as well as the local photon density of states (i.e., $LDOS({\bf r},\omega) \propto
{\rm Trace}\{{\rm Im}[\overline{\bf G}({\bf r},{\bf r};\omega)]\}$).
For convenience, we partition the Green function into contributions from the bound waveguide mode, radiation modes above the light line, and other modes as
$	\G(\R,\R';\omega) = \G[B](\R,\R';\omega) + \G[R](\R,\R';\omega) + \G[O](\R,\R';\omega)$.
The bound mode Green function is given analytically from properties of the bound mode \cite{Patterson:MScE,Hughes:2005}:
$\G[B](\R,\R';\omega) = i \frac{a \omega}{2 v_g}
		[  \e_{k}(\R) \otimes \e_{k}^*(\R')  e^{ik(x-x')}  \Theta(x-x')
			 + \e_{k}^*(\R) \otimes \e_{k}(\R')$ $ e^{ik(x'-x)}  \Theta(x'-x) ]$,
where $v_g$ is the group velocity,
 $\e_k(\R)$ is the 
 Bloch mode electric field normalized by $\int_\mathrm{cell} \id \R \, \varepsilon(\R) \, |\e_k(\R)|^2  = 1$, $\otimes$ is a tensor product, $\e_{-k}(\R) = \e_k^*(\R)$, and $\Theta(x)$ is the Heaviside step function.
%
For our calculations below, we use a W1 PC 
waveguide formed from a semiconductor membrane by omitting a row of holes in a two-dimensional triangular array of holes. The lattice pitch is $a=480\nm$, the hole radius $r=95\nm$ and the slab thickness is $h=160\nm$. The
dispersion of the 
{\em ideal} 
waveguide mode is shown in Fig.~\ref{fig:bandstructure}(a). The electric field 
Bloch 
mode distribution is shown on symmetry planes in the unit cell in Fig.~\ref{fig:bandstructure}(b) for a wave vector near the mode edge.

The disorder fluctuations of interest must closely correspond to real images of fabricated PC waveguides~\cite{Skorobogatiy:2005}, and thus we consider disorder in PC slab structures that is dominated by perturbations of the perimeter of the holes, an example of which
 is shown
 in Fig.~\ref{fig:RoughHole}(c); this is also consistent with previous models used to successfully fit experiments \cite{Kuramochi:2005,Patterson:2009b}. We take the radial perturbation $\Delta r$ to be a Gaussian random variable with a mean of zero and a standard deviation of $\sigma$. Two radial perturbations are correlated by
$	\langle \Delta r(\phi_i) \Delta r(\phi_j') \rangle = \sigma^2 \, e^{-r|\phi_i - \phi_j'|/l_p} \, \delta_{i,j}$,
where the subscript indexes the holes, $\phi_i$ is the angular position of the point measured about the centre of the hole, $r$ is the hole radius, and $l_p$ is the correlation length measured around the circumference. The change in dielectric constant can be written as
$
	\Delta \varepsilon(\R) = (\varepsilon_2 - \varepsilon_1) \, \Delta r \, \delta(\sqrt{x_i^2 + y_i^2} - r) \, \Theta(h/2 - |z|)
$,
where $x_i$ and $y_i$ are the components of $\R$ measured from the centre of the hole, and the Heaviside step function restricts the disorder to a slab of thickness $h$. In this analysis, below, we  used representative disorder parameters of $\sigma = 3\nm$ and $l_p = 40\nm$, which are typical for fabricated  samples.

The failure of the weak contrast model, namely ${\bf P}_{\rm dis}= \Delta\varepsilon {\bf E}_0$ is most obvious by considering a small dielectric sphere introduced in a homogeneous background with dielectric constant $\varepsilon_2$. It is well known that the weak contrast polarizability $\Delta \varepsilon$ must be replaced with the correct polarizability $3 \Delta \varepsilon / (3 \varepsilon_2 + \Delta\varepsilon)$ (as can easily be proven from Eq.~(1) \cite{Ramunno:2009}). Due to local field effects, the macroscopic scattering depends on the microscopic geometry of the scatter, and this is also true for disordered waveguides. This issue has been partly investigated by \citet{Johnson:2005}, who give the correct disorder polarization density due to a disorder element at $\R'$ as ($\omega$ is implicit)
\begin{equation}
	\mathbf P(\R) = \left(\frac{\varepsilon_1 + \varepsilon_2}{2} \alpha_\parallel \mathbf E_\parallel(\R)
		+ \varepsilon(\R) \gamma_\bot \mathbf D_\bot(\R) \right)\! \Delta V  \delta(\R - \R'),
	\label{eqn:polarization_Johnson}
\end{equation}
where $\alpha_\parallel$ and $\gamma_\bot$ are polarizabilities for the disorder element, $\varepsilon(\R)$ takes a different value depending on which side of the interface $\R'$ is located, and $\Delta V$ is the volume of the disorder element. Due to the $\varepsilon(\R)$ factor, $\mathbf E^*(\R) \cdot \mathbf P(\R)$ will have terms proportional to $|\mathbf E_\parallel(\R)|^2$ and $|\mathbf D_\bot(\R)|^2$ and these fields are well defined at a dielectric interface. This improved disorder model hides a great deal of complexity in the polarizabilities $\alpha_\parallel$ and $\gamma_\bot$. The polarizabilities are different for a positive bump ($\varepsilon_1$ extending into $\varepsilon_2$, $\alpha_\parallel = \alpha_\parallel^+$) and a negative bump ($\alpha_\parallel^-$). Further, in general, the polarizabilities are asymmetric so that $\alpha_\parallel^+ \ne -\alpha_\parallel^-$. Finally, the exact polarizabilities depend on the precise geometry of the disorder element and must be calculated numerically. It is thus no surprise that only a single isolated bump was treated in Ref.~\citenum{Johnson:2005}.

One method to simplify the treatment of the polarizabilities is to assume a structure for the disorder where the polarizability is known. In Ref.~\citenum{Wang:2008}, a polarization density of the form
$	\mathbf P(\R) = \Delta\varepsilon(\R) \left(\mathbf E_\parallel(\R)
		+ \varepsilon(\R) \frac{\mathbf D_\bot(\R)}{\varepsilon_1 \varepsilon_2}
		\right) \delta(\R - \R')$
was used, namely the {\em slowly-varying surface approximation}. This takes care of the issue of the $\mathbf E(\R)$ being ill-defined at the interface but it has some subtle and questionable assumptions. Implicit in the derivation of this term is the assumption that the disordered surface is smooth; effectively that the hole radius changes but remains nearly circular. As a model for the disorder in real fabricated samples, this is rather suspect, and it is known that the interface fluctuations vary rapidly (cf.\ Fig.~\ref{fig:RoughHole}(c) and Ref.~\citenum{Skorobogatiy:2005}).
Our solution to this general problem is to use a representative model for the polarizability, namely Eq.~\ref{eqn:polarization_Johnson} with the numerically calculated polarizabilities for a cylindrical bump \cite{Johnson:2005}, in combination with our numerically generated disordered profile. We use this straightforward and qualitative approach to provide essential physics insight into optical scattering phenomena that may occur when the polarizability is correctly treated, which, as mentioned previously, is lacking in all previous disorder models.

\begin{figure}
	\centering
	\includegraphics{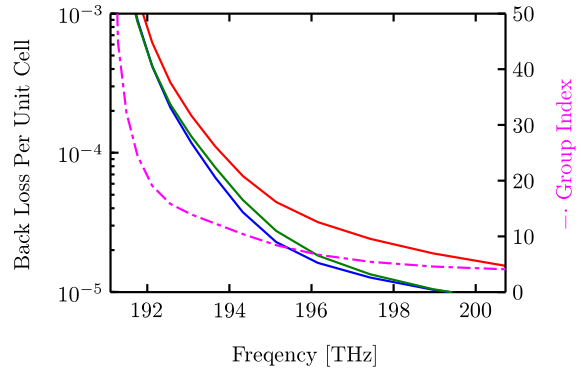}
\vspace{-0.0cm}
	\caption{\label{fig:incoherentloss} (Color online) Incoherent averaged back scatter loss for a single unit cell using the weak contrast (blue), smooth surface (green) and cylindrical bump (red) polarization density models. 
For reference, the group index (magenta, dot-dash) is shown on the right scale.}
\end{figure}

It is easiest to compare these polarization density models above using our incoherent scattering calculation where the band-edge resonances are suppressed. The incoherent average power loss, which is dominated by backscatter loss for slow light PC waveguides, is \cite{Hughes:2005}
\begin{align}
	\langle \alpha_\mathrm{back} \rangle \! = \! \left(\frac{a \omega}{2 v_g}\right)^2 \!\!\! \iint \! \id \R  \id \R'
\left\langle
		\left[\mathbf E(\R) \! \cdot \! {\mathbf P(\R)} \right]
		\left[\mathbf E^*(\R') \! \cdot \! {\mathbf P^*(\R')} \right]
		\right\rangle,
\end{align}
where the integration is performed over a single unit cell and $\mathbf E(\R)={\bf e}_k(\R) e^{ikx}$.
The incoherent averaged backscatter loss is plotted in Fig.~\ref{fig:incoherentloss} for the weak contrast (blue), smooth surface (green), and cylindrical bump (red) models. Although the exact loss predictions differ, all the models predict similar loss trends and their magnitudes are similar. For the roughness statistics typical of PC waveguide systems, we find reasonable agreement between the backscatter loss predictions of the three models, though the smooth surface and weak contrast models underestimate the scattering loss. 
But this by itself, although an improvement, is not a drastic revision of previous disorder models.

\
%
As discussed previously, one of the effects neglected when one ignores the $\G[O](\R,\R';\omega)$ contribution of the Green function is disorder-induced frequency shifts. This--hitherto--theoretically unknown  phenomenon for PC waveguides is closely related to the selection of a suitable polarization model, as the predicted frequency shift is sensitive to this term. The first-order {\em mean frequency shift} due to disorder is given as\cite{Sakoda:2005}
\begin{equation}
	\langle \Delta \omega \rangle = -\frac{\omega}{2} \int \id \R \left\langle \mathbf E^*(\R) \cdot {\mathbf P(\R)} \right\rangle
	\label{eqn:domega}
\end{equation}
with $\mathbf E(\R)$ normalized as before.
%
The frequency shift is usually taken to be zero for zero-mean surface perturbations, but as shown previously for a PC cavity \cite{Ramunno:2009}, correctly treating local field effects yields a non-zero first order frequency shift. Considering the above polarization densities, both the weak contrast and smooth surface models predict $\langle \Delta \omega \rangle = 0$ due to the symmetry in the polarization for positive and negative bumps. The cylindrical bump polarizability predicts a non-zero $\langle \Delta \omega \rangle$, but the details differ from Ref.~\citenum{Ramunno:2009} (the simpler PC cavity case) due to differing disorder models; for the waveguide, we are also dealing with a continuous mode rather than a discrete resonance. When using the polarization density of Eq.~\ref{eqn:polarization_Johnson}, care must be taken when evaluating the expectation value since the value of the polarizabilities depends on the direction of the bump. In addition to the mean frequency shift, which in general may or may not be zero, the RMS frequency shift $\langle \Delta \omega \rangle_{RMS} = \sqrt{\langle\Delta\omega^2\rangle}$ is certainly not. It is calculated in a similar way to the corrected backscatter loss, from
\begin{equation}
	\langle \Delta \omega^2 \rangle = \frac{\omega^2}{4} \iint \id \R \, \id \R' \left\langle
		\left[\mathbf E^*(\R) \cdot {\mathbf P(\R)} \right]
		\left[\mathbf E^*(\R') \cdot {\mathbf P(\R')} \right]
		\right\rangle.
\end{equation}
We stress that neither of these two frequency shifts have been predicted nor calculated for PC waveguides, yet clearly they are just as important, if not more so, as computing the disorder-induced power loss.
\begin{figure}
	\centering
	\includegraphics{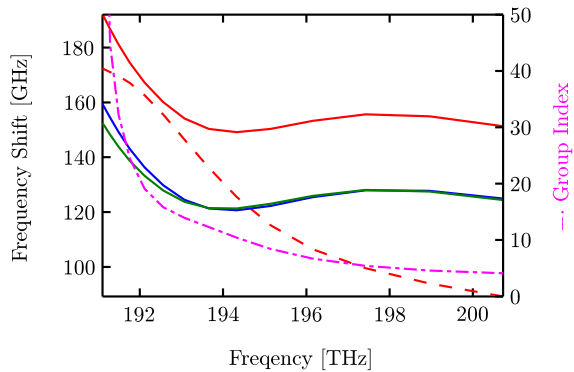}
\vspace{-0.0cm}
	\caption{\label{fig:freqshift} (Color online) Mean frequency shift (dashed) and RMS frequency shift (solid) using the weak contrast (blue), smooth surface (green) and cylindrical bump (red) polarization density models. Only the cylindrical bump model has a non-zero mean frequency shift. For reference, the group index (magenta, dot-dash) is shown on the right scale.}
\end{figure}
%
Figure \ref{fig:freqshift} plots the mean (dashed) and RMS (solid) frequency shifts for the weak contrast (blue), smooth surface (green) and cylindrical bump (red) polarization density models. The weak contrast and smooth surface models---{\em incorrectly}---predict zero mean frequency shifts  and thus are not shown. The disorder-induced frequency shifts are particularly important for understanding experimental transmission spectra. Typically, very near the band edge, there will be an abrupt drop in the transmission associated with a local frequency shift causing the band to be shifted such that the injected frequency is below the new band edge. As can be seen from the RMS frequency shifts, this will be an issue regardless of the polarization model, but can be exasperated by a non-zero mean frequency shift.  The inset to Fig.~1(a) depicts the computed band structure with disorder\cite{note2} (grey shaded band), showing that getting too close to the mode edge is actually impossible. Qualitatively, for waveguide mode frequencies within a few standard deviations of the mode edge (slow light regime), we expect that somewhere along the waveguide, total reflection will inevitably occur.

These results are  
directly 
relevant to
a wide range of enhanced light-matter interaction physics that
occurs in nanophotonic waveguides. 
Apart from the consistent observation of a sudden, dramatic reduction in transmission
in PC waveguide near the mode edge (underestimated by current theories), we
 cite two other examples: 
(i) observation of the enhanced spontaneous emission with single photon emitters (quantum dots) near the mode edge indeed observe
a broadened and substantially reduced density of states than
is expected from the disorder-free band structure \cite{lodahl:2008};
(ii)  recent work
by Morichetti {\em et al.} \cite{Morichetti:PRL2010} demonstrate that backscattering is also
one of the most severe limiting factors in state-of-the art silicon on insulator nanowires, and our
conclusions are directly applicable and supportive of these measurements as well: broadening and
band structure restructuring near the mode edge will, inevitably, result in dramatic backscattering.


In summary, we have described a theory of disorder-induced scattering to include the influence of local field effects and high-index-contrast perturbations. Our calculations are shown to increase the predicted waveguide losses and result in significant and surprising (asymmetric) disorder-induced resonance shifts. The band structure broadening (schematically shown in Fig.~\ref{fig:bandstructure}(a)) offers fresh and important insights into the
fabrication limits of slow light propagation in PC waveguides and should serve as a further warning that propagation modes near the mode edge will have an increasingly better chance of being completely reflected. Our predictions are 
consistent with the wide range of experiments in the literature.

This work was supported by the National Sciences and Engineering Research Council of Canada,
and the Canadian Foundation for Innovation.





\begin{thebibliography}{22}
\bibitem[{not({\natexlab{a}})}]{note1}
\bibinfo{note}{For conciseness of notation, the polarization densities used in
  this paper have $\varepsilon_0$ factored out.}
\bibitem[{not({\natexlab{b}})}]{note2}
\bibinfo{note}{The standard deviation of the frequency shift about the new mean
  value is defined as $\sqrt{\langle \Delta \omega^2 \rangle - \langle \Delta
  \omega \rangle^2}$.}
\end{thebibliography}

\end{document}